\documentstyle[a4wide,epsf,11pt,titlepage]{article}

\newcounter{nref}
\newcommand{\bbib}{%
  \renewcommand{\refname}{\large\bf References}%
  \begin{thebibliography}{99}%
  \setcounter{enumiv}{\thenref}}
\newcommand{\ebib}{%
  \end{thebibliography}%
  \setcounter{nref}{\arabic{enumiv}}}
\newcommand{\head}[3]{%
  \setcounter{nref}{0}%
  \thispagestyle{empty}%
  \section*{\LARGE\bf #1}%
  \stepcounter{section}%
  \addcontentsline{toc}{section}{#1}%
  \large\itshape%
  #2\\\vspace{0.1pt}\\%
  #3%
  \normalsize\upshape%
  \bigskip}

\begin{document}

%----------------------------------------------------------------------

\head{Disturbance Ecology from Nearby Supernovae}
     {D.H.\ Hartmann$^1$, K. Kretschmer$^2$, and R.\ Diehl$^2$}
     {$^1$ Clemson University, Department of Physics \& Astronomy, Clemson, SC 29634-0978\\
     $^2$ Max-Planck Institut f\"ur Extraterrestrische Physik, PO Box 1312, 85741 Garching}

\section*{The Hammer of God}
``Kali 2 entered the atmosphere just before sunrise, a hundred kilometers
above Hawaii. Instantly, the gigantic fireball brought a false dawn to the 
Pacific, awakening the wildlife on its myriad islands. But few humans; not
many were asleep this night of nights, except those who had sought the 
oblivion of drugs'' \cite{hartmann.clarke}.

This grim description of an asteroid impact by Arthur C. Clarke was inspired
by the famous paper ``Extraterrestrial Cause for the Cretaceous-Tertiary Extinction''
by Nobel Laureate Luis Alvarez and his geologist son, Walter Alvarez
\cite{hartmann.alvarez}. That great extinction episodes in geological
history are tightly connected to asteroid/comet impacts is now firmly
established \cite{hartmann.becker}\cite{hartmann.Trex}. Whether impact events are
(quasi)periodic or random is not yet clear, and it is also not yet established
if some external agent (e.g., a hypothetical Nemesis companion star of the Sun
or modulated Oort cloud pertubations via solar oscillations in the Galactic disk;
\cite{hartmann.muller}\cite{hartmann.raup})
is required to explain the extinction record. While the idea of repeated blows by
some ``Hammer of God'' has found widespread acceptance, a paradigm shift is underway
with regards to the way we think about such catastrophes. Was the extinction of
the dinosaurs a bad day in our history? From the point of view of the dinosaurs
it sure counts as a bad day, but from our present-day perspective we can appreciate
how this impact promoted evolutionary changes that benefited humans and other species.

\section*{Disturbance Ecology on Grand Scales}
Ecology is the study of the interrelationships between organisms and
their environment. In the past interactions between predators and prey
and between herbivors and plants held the preeminent status among all
ecological processes, with the environment merely providing the stage on 
which the ecological play takes place. This view has changed dramatically in
recent decades. The ``New Ecology'' endowes the environment with an
active role. Fires, floods, hurricanes, tornadoes, tsunamis, land-slides,
vulcanic eruptions, storms, and lightning strikes are all key agents of 
the new ``disturbance ecology'' \cite{hartmann.reice}\cite{hartmann.court}.
The destructive power of nature is now viewed as essential for maintaining
biodiversity. This new persepective can be 
expanded by considering that changes in the solar system environment can
directly affect the Earth, and that such changes can be driven by gentle
or catastrophic events in the solar neighborhood. Long term evolution of
the Sun will steadily increase the equilibrium temperatures of the planets
on time scales of billions of years (melting the ice caps and eventually boiling 
the oceans and evaporating the atmosphere), while solar flares occur frequently
and occasionally can be so energetic that global planetary atmospheres
and the interplanetary space environment become severely polluted with energetic
particles that can significantly alter atmospheric chemistry \cite{hartmann.chap}
\cite{hartmann.schaefer}\cite{hartmann.hoyt}. 
Other dangers lurk just outside the solar system.

As the Sun moves around the Galactic center in a nearly circular orbit the Galactic
environment in general changes slowly, but ``sudden'' (10$^{3-4}$ years) 
modifications of the planetary environments can take place when the Sun encounters 
local density fluctuations in the interstellar medium \cite{hartmann.verschuur}
\cite{hartmann.ZF99} or local field stars
\cite{hartmann.gs}. The latter situation can disturb the
Oort cloud, increasing the cometary impact rates in the solar system, while the 
former situation can modulate the properties of the heliosphere and thus the local
cosmic ray environment, with potential consequences for planetary climates. Still
more damaging could be encounters with very massive stars, black holes, novae, super-
novae, and hypernovae (aka Gamma Ray Bursts, GRBs).   

\section*{Exposure to Nearby Gamma Ray Bursts}

The largest energy releases in the Universe due to supernovae and gamma-ray
bursts (\cite{hartmann.stsci} and \cite{hartmann.wheeler} discuss both phenomena 
and their possible connections) do take place with somewhat predictable rates. A
core collapse supernova occurs about once every second in the Universe. The 
Milky Way's present star formation rate is a few solar masses per year, which 
for a Salpeter-like IMF translates into a Galactic supernova rate of about once 
per century. The GRB rate is of order 1 day$^{-1}$, and we know that bursts
are truly cosmological (typical redshifts of order unity, with a present record 
at z $\sim$ 4.5
for GRB000131). Afterglow observations point towards a direct link between
GRBs and supernovae so that bursts are also expected to occur in the Milky Way,
although with a lower rate. The relative frequency of SNe and GRBs depends on the
uncertain beaming pattern of bursts. The observations imply significant beaming 
corrections, but suggest that GRB rates are still much lower than supernova rates.
Long-duration GRBs and SNe are expected to have similar spatial distributions. 
The amount of energy
released into the surrounding medium is of order 1 foe (10$^{51}$ ergs) in both
cases, but GRBs focus energy into narrow jets while SNe impact their environments
$\sim$isotropically.

Hartmann \cite{hartmann.dino} and Thorsett \cite{hartmann.thorsett} considered
the possibility that a GRB was responsible for the demise of the dinosaurs and
perhaps other mass extinctions. Thorsett studied in detail the effect of the 
gamma-ray flux on Earth's atmosphere, especially the reduction of ozone and the
subsequent effects of enhanced solar UV radiation on the surface. Thorsett 
concluded that the critical distance was D$_{\rm crit}$ $\sim$ 1 kpc,
and estimated a rate of events closer than this distance of about 1/10$^8$ yrs.
In 1998 Dar, Laor, and Shaviv \cite{hartmann.dls} investigated the effects of
cosmic ray exposure for planet Earth located inside the jet of a galactic GRB.
These authors concluded that ozone destruction and radioactive pollution of the
Earth could have contributed to mass extinctions, and that biological effects
due to ionizing radiation could also have caused the fast appearance of new species.
More recently Dar $\&$ DeRujula \cite{hartmann.dd} considered the potential
dangers from Eta Carina as a future nearby GRB source (concluding that the 
direction of the possible jet is not aiming at us), and determined that
jet-sterilization per life-supporting planet could happen as often as every
100 million years. Such a high rate might provide an answer to Fermi's famous
question about alien visitors: ``Where are they?". Depressing thoughts along 
these lines were the theme of the recent NOVA show "Death Star" 
\cite{hartmann.nova}; nothing seems to sell better than sex and doomsday predictions
\cite{hartmann.doom}. 

A less sensational approach to the possible biological
effects of nearby bursts was recently presented by Scalo and Wheeler 
\cite{hartmann.sw}, who investigated in great detail the effects of
ionizing GRB radiation on DNA alterations and chemical modifications
of the atmosphere. The emphasis of this work is thus not on catastrophes
but smaller (more frequent and perhaps cumulative) events that can 
alter evolution through enhanced mutagenesis. Scalo and Wheeler estimate 
the ``critical '' X-$\gamma$-fluence required for a significant increase
in DNA alteration, F$_{\rm crit}$ = 5~10$^5$ ergs cm$^{-2}$.
From observed GRB spectral properties the sphere of influence of a galactic 
GRB is found to be $\sim$ 10 kpc, so that many galactic GRBs can have a 
potential impact on Earth's ecology. Scalo and Wheeler estimate that at 
least one thousand biologically significant Galactic GRB irradiations
should have occurred stochastically during the 4.4 Gyr history of life
on Earth. The direct effect of one burst (GRB830801) on the ionosphere was recorded
via changes in the VLF radio transmissions between
distant sites \cite{hartmann.inan}, and this event most likely occured at a 
cosmological distance. Although GRBs are very luminous, their short duration 
provides a moderating factor
in comparison to other potential sources of ionizing radiation (massive
stars, supernovae, etc), and causes only one half of the Earth to be affected
in any given exposure. 

\section*{Exposure to Nearby Supernovae}

The rate of nearby Galactic supernovae is a rather uncertain quantity, and
estimates date back to the sixties \cite{hartmann.shklovsky}. The impact
of a close supernova on the biosphere was considered in the seventies
\cite{hartmann.ruderman}\cite{hartmann.hunt}\cite{hartmann.whitten} and 
subsequent studies zoomed in on particular supernovae (e.g., Vela 
\cite{hartmann.brakenridge}, and Geminga \cite{hartmann.gehrels}) or nearby
star forming regions (Sco-Cen \cite{hartmann.benitez}). The potential 
identification of such events through specific isotopic anaomalies
(e.g., $^{60}$Fe) was pointed out by Ellis and collaborators
\cite{hartmann.efs}\cite{hartmann.fe}. A recent search for $^{60}$Fe anomalies
in deep-ocean ferromanganese crust samples was sucessful \cite{hartmann.knie}
and estimates of the distance and epoch of the putative supernova give 
D $\le$ 30 pc and age $\le$ 5 Myrs \cite{hartmann.fe}, based on standard yields
of M$_{\rm ej}$($^{60}$Fe) $\sim$ 10$^{-5}$ M$_\odot$ \cite{hartmann.timmes}.
The decay of $^{60}$Fe (mean life $\sim$ 2 Myrs) produces gamma-ray lines
at 1.17 MeV and 1.33 MeV. The diffuse glow of the galactic disk from the many
supernovae that occur during a few mean life times may soon be detected with 
INTEGRAL, an ESA mission to be launched later this year \cite{hartmann.integral}
\cite{hartmann.vos}. Tracing the orbits of the Sun and nearby
stars back in time Benitez et al. \cite{hartmann.benitez} 
find that massive stars born in the Sco-Cen association 
produced supernovae at the right time and distance to be responsible for the
$^{60}$Fe observed in the ocean crust. According to these authors $\sim$ 2 Myrs
ago a ``Sco-Cen supernova'' as close as D $\sim$ 40 pc could have done serious 
damage to the Earth' ozone layer, provoking or contributing to the Pliocene-
Pleistocene boundary marine extinction. A very recent (or perhaps near future)
supernova impact may be associated with ``Vela-Junior'' \cite{hartmann.aschenbach},
for which $^{44}$Ti gamma-ray line observations \cite{hartmann.iyudin} were used
to argue for a distance of D $\sim$ 150 pc and an age of t $\sim$ 700 yrs.

While dramatic events such as the demise of the dinosaurs and other mass
extinctions may in some cases be linked to nearby supernovae \cite{hartmann.chap}, 
more moderate alterations of evolutionary paths through the enhanced UV exposure 
were only recently considered in detail \cite{hartmann.sww}. Core collapse 
supernovae launch a strong shock that propagates through the star for a
few hours, depending on the progenitor size and mass and ejecta energy. 
The shock-breakout leads to a short UV-flash \cite{hartmann.dopita}
\cite{hartmann.ensman}\cite{hartmann.blinnikov} that delivers of order 
10$^{47}$ ergs in the 200-300 nm UV band. Scalo et al estimate that a 
``critical" fluence for enhanced mutagenesis in this crucial energy
window is F$_{\rm crit}$ $\sim$ 600 ergs/cm$^2$. Taking into account 
absorption in the atmosphere the relevant value at the top of
the atmosphere is about three times as large. The critical distance for
shock breakout UV-flashes to affect the biosphere is thus of order 1 kpc.
The effect of galactic extinction must also be taken into account, as
emphasized by Scalo et al. In the
UV band considered here the average differential midplane UV-extinction
is $\partial_r$A$_{UV}$ $\sim$ 5 mag kpc$^{-1}$. In addition, one should 
consider the spatial distribution of the 
absorbing dust relative to the spatial distribution of supernovae.    

\begin{figure}[ht]
  \centerline{\epsfxsize=0.6\textwidth\epsffile{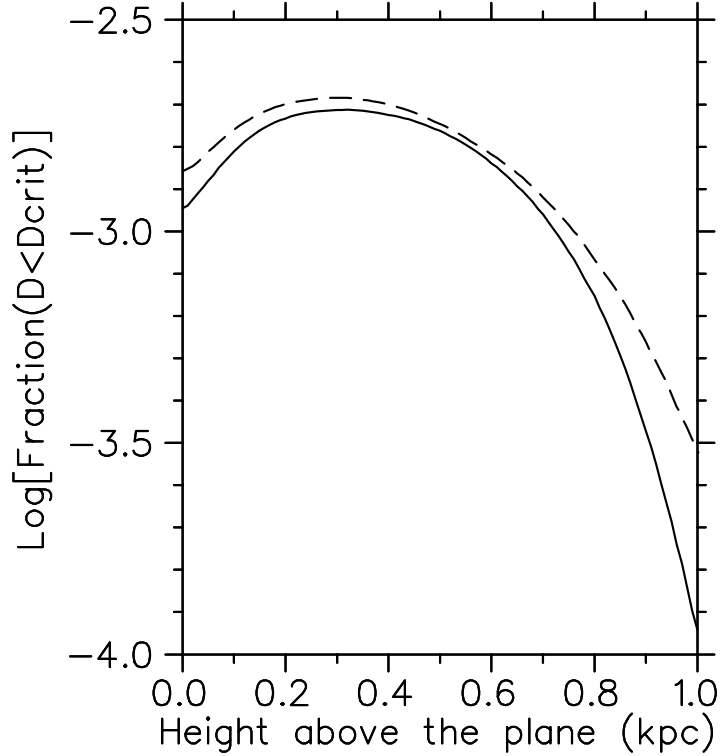}}
  \caption{The fraction of Galactic Supernovae closer than the
   critical distance (see text) as a function
   of position above the midplane. The solid line assumes
   the same vertical profile for dust and supernovae, while the 
   dashed curve is for a supernova distribution with doubled scale height.} 
  \label{hartmann.fig1}
\end{figure}

What is the frequency of supernovae that generate UV fluences at Earth
that exceed the critical value? A galaxy-wide star formation rate
of a few solar masses per year is associated with a supernova rate of
a few events per century (many papers have been written about the value
of ``few'' in both of these quantities). If the fraction of events closer
than D$_{\rm crit}$ is f, the time between ``events that cause a disturbance''
is $\Delta$T $\sim$ 100/{\rm few}~f$^{-1}$ ${\rm years}$, where ``few'' is
about 1$-$3 \cite{hartmann.tam}. 
We find that f is
of order 10$^{-3}$, so that dangerous UV-flashes from nearby supernovae
occur as often as every $\sim$100,000 years. Scalo et al. provide an estimate
of this time scale as a function of fluence and peak flux. 
  
We now consider a modification of these rate estimates due to the solar motion
in the Galaxy. The Sun is currently located $\sim$ 10 pc above the midplane 
(this number is uncertain, but everyone agrees that it is not zero 
\cite{hartmann.min}\cite{hartmann.ham}\cite{hartmann.ng}\cite{hartmann.coh}),
but its perpendicular velocity relative to the local standard of rest W =
7.2 $\pm$ 0.4 km/s \cite{hartmann.bm} will eventually carry us to a maximal
height H $\sim$ 100 pc (depending on the galactic mass model) above
the plane. The sun, like most stars, will spend most of its time above the
Galactic dust layer, so that this effect should be included in estimates
of the extinction to supernovae. The figure shows the variation of f with 
height above the Galactic plane. For the Sun the change in $\Delta$
T is less than a factor two, but for planets around high-velocity stars the
effect can be significant (making damaging exposure ten times less frequent
if vertical motion carries the star to z $\sim$ 1 kpc above the plane. The
effect seen in the figure is due to the exposure asymmetry (events
above/below the plane) for small heights but simply the result of larger
distances from supernovae as the star moves away from the disk. For heights
well in excess of 1 kpc none of the galactic supernovae is ever close enough
to have an effect. As the stars oscillate in the Milky Way with periods of
order 50$-$100 Myrs plane passages would lead to increased exposure, but less time 
is spend
in the ``critical midplane zone''. A Monte Carlo simulation using these constraints
could evaluate the time-averaged UV-exposure as a function of position and kinematics
in the Galaxy.

We also investigated the effects of a larger scale height of supernovae
relative to the dust (dashed line in the figure). The standard model assumes
an exponential scale height of H$_z$ = 60 pc for both components, while the 
model with larger extent of the supernova layer assumes H$_z$ = 120 pc. The
effect of ``puffing up'' the SN distribution is reduced extinction and thus
a slightly larger fraction of critical events, but the effect is obviously not 
very strong. 
The radial distribution of supernovae probably is a more critical parameter. 
It is however rather uncertain and we 
describe it by a gaussian function, centered
at R = 4 kpc, the location of the well known molecular ring \cite{hartmann.bm}.
If the width of this gaussian drops below 3 kpc, the fraction of ``deadly SN''
decreases rapidly. Clearly, a delta function at R = 4 kpc generates no
SN close enough to damage the biosphere. If the gaussian width exceeds
3 kpc, the fraction saturates as the local SN density profile becomes flat
(as was assumed by Scalo et al.). While the time between SN that are ``too close
for comfort'' depends on many properties of the Galaxy, the solar orbit, the 
shock breakout properties, and assumptions about what constitutes a 
biologically significant exposure, the perhaps surprising fact remains that
close calls happened many times along our path around the galactic center.

\section*{Conclusions and Outlook}

Can a nearby GRB or Supernova ruin your day ? You bet! Does this happen
frequently? Yes and no, depending on your point of view. On a present-day
human history time scale the answer is no (fortunately), but on a long-term-
history-of-life time scale the answer is yes. Over the past billions of
years there must have been a large number of close encounters of the 
undesired kind - UV jolts from supernovae and GRBs, and longer
exposures to passing luminous and hot stars. Disturbance ecology is driven
by events taking place on Earth and in the immediate Galactic environment
of the solar system. Mutagenesis can be stimulated by exposure to external 
UV sources, and despite the long time intervals between events (ten thousand
years to perhaps millions of years) these ``rare'' events may have domiated
the path of evolution on Earth as well as other inhabited planetary systems
throughout the Galaxy. The killing of the dinosaurs was indeed spectacular,
but perhaps more impressive is the realization that the evolution of life
throughout the Universe is so closely intertwined with the violent history
of the Universe. Once every second a supernova and once every minute a GRB
sends a UV flash into some galactic environment, jolting mildly or vehemently
life in its path. It happened many times
in Earth's history, and it will happen again. We hope that the next nearby
jolt in our path contributes to a better global future for most lifeforms 
on Earth. 

\subsection*{Acknowledgements}

With pleasure we thank the organizers 
and the staff for outstanding hospitality   
and a very stimulating meeting.

%----------------------------------------------------------------------

\end{document}